\documentclass[prb,preprint,unsortedaddress]{revtex4-1}

\usepackage{amsmath}  
\usepackage{amsfonts} 
\usepackage{graphicx} 

\begin{document}


\centerline{\bf Complete positivity on the subsystems level}

\centerline{ M. Arsenijevi\' c$^1$, J. Jekni\' c-Dugi\' c$^2$, M. Dugi\' c$^1$}

$^1${University of Kragujevac, Faculty of Science, 34000 Kragujevac, Serbia}

$^2${University of Ni\v s, Faculty of Science and Mathematics, 18000 Ni\v s, Serbia}


\date{\today}

\bigskip

\noindent {\bf Abstract} We provide a conceptually clear and technically simple presentation of certain subtleties of the concept of complete positivity of the quantum dynamical maps. The presentation is performed by addressing complete positivity of dynamics of certain subsystems of an open composite system, which is  subject of a completely positive map.
We prove that
every subsystem of a composite open system can be subject of a completely positive dynamics if and only if the initial state of the composite open system is tensor-product of the
initial states of the subsystems. A general algorithm for obtaining the Kraus form for a subsystem's dynamical map is designed for the finite-dimensional  systems. As an illustrative example we consider a pair of mutually interacting qubits.

\bigskip


\bigskip

\maketitle 

\noindent {\bf 1 Introduction} 

\bigskip

\noindent The open quantum systems theory [1, 2] is at the core of the modern quantum theory and applications, ranging from the foundational issues [1-8]
and quantum optics [1, 9] to the cosmological issues [10, 11] and the diverse applications, e.g., in quantum technology [12], materials science [13], chemistry and nanotechnology [14-17] as well as in the nascent field of quantum thermodynamics [18, 19].

For the physical reasons, complete positivity [1, 2, 20] of the open system's dynamics is assumed to be without viable alternative. However, complete positivity (CP) of the open system's dynamics (dynamical map) does {\it not} necessarily follow from the {\it microscopic} model of the composite system "open system + environment ($S+E$)" [1, 2, 21, 22]. That is, while the  assumption of the unitary dynamics for the
$S+E$ system is used, the subtle points regarding the model of the environment and the coupling in the $S+E$ system may lead to a non-CP dynamics for the open system [1, 2, 23, 24]. In addition to this, the very concept of complete positivity may require a careful redefinition [25-28].

In this paper, we provide a {\it simple and general}  presentation of the subtleties [1, 2, 20-28] regarding the concept of complete positivity.
The presentation follows directly from an answer to the following question: assuming that an open composite system is subject of a CP dynamics, under which conditions the  subsystems' dynamics can also be completely positive?
 As an illustration of the general statements, we use a model of a pair of interacting qubits [29] that illustrates the technical finesse of general interest for the  bipartitions of the composite quantum systems. The presentation is technically as well as conceptually simple in the context of  "quantum structures", i.e. in the context of the different decompositions of a composite system into subsystems [5, 6, 30, 31]. That is, with some general rules regarding the quantum structures, the conceptual subtleties regarding complete positivity become rather transparent. As a benefit of answering the above-posed question, we provide an algorithm for deriving the Kraus forms for the subsystems CP dynamics.

In Section 2 we emphasize the different possible contents of the concept of "completely positive dynamics" that can be found in the literature. In Section 3 we emphasize the concepts and nomenclature of the so-called quantum structures (notably bipartitions) of the composite quantum systems. In Section 4, we provide the main results of this paper leading to a formulation of an algorithm for deriving the Kraus operators for the subsystems of an open system. Section 5 provides a clear yet general illustration of the delicate concept of the domain-dependent complete positivity. In Section 6 we provide an application of the general results obtained in Section 4. Section 7 is discussion and we conclude in Section 8.

\bigskip

\noindent {\bf 2 Complete Positivity}

\bigskip

\noindent By definition, the open system's state $\rho_S(t)$  is  ''statistical operator'' (''density matrix''), which is a Hermitian ($\rho_S^{\dag}=\rho_S$), unit trace ($tr_S\rho_S=1$) and semidefinite-positive operator, whose
eigenvalues are  nonnegative real numbers. Statistical operators for a system $S$ constitute a Banach vector state space, $\mathcal{B}_S$. Dynamics of the open system $S$ is described by a dynamical map $\Phi_{(t,t_{\circ})}$, which transforms the state in an initial $t_{\circ}$ to a final $t\ge t_{\circ}$ instant of time: $\rho_S(t) = \Phi_{(t,t_{\circ})} \rho_S(t_{\circ.})$.

We are interested  in the dynamical maps satisfying the following conditions:

\noindent (i) Both the initial $\rho_S(t_{\circ})$ and the final $\rho_S(t)$ state are elements of the system's state-space  $\mathcal{B}_S$,

\noindent (ii) The map preserves the trace, i.e. $tr_S\rho_S(t) = 1, \forall{t\ge t_{\circ}}$,

\noindent (iii) Domain of the map is the {\it whole} state space, $\mathcal{B}_S$, i.e. every possible state $\rho_S\in \mathcal{B}_S$.

It is natural to assume that dynamics of the open system $S$ should not be influenced by another system $A$, which the $S$ system had never interacted with in the past.
This is the original requirement of complete positivity  for the dynamical map $\Phi_{(t,t_{\circ})}$ [1, 2, 20]. Due to the Kraus theorem [20],
complete positivity of a dynamical map is {\it equivalent} with the possibility to present the map in the form:

\begin{equation}
\Phi_{(t,t_{\circ})} \rho_S(t_{\circ}) = \rho_S(t) = \sum_k K_{Sk}(t) \rho_S(t_{\circ}) K^{\dag}_{Sk}(t).
\end{equation}

 A CP map eq.(1) that satisfies the above condition (ii) of trace preservation also fulfills the so-called completeness condition:

\begin{equation}
\sum_k K^{\dag}_{Sk}(t) K_{Sk}(t) = \mathcal{I}_S,
\end{equation}

\noindent where $\mathcal{I}_S$ represents the identity map in $\mathcal{B}_S$. The maps satisfying eqs.(1) and (2) are often called {\it completely positive and trace preserving}  maps.

 In order to ease the presentation, here and further on, we assume the following definition of CP:

\noindent {\bf Definition 1} By {\it completely positive dynamical maps} we assume the  maps that can be written in the form of eq.(1) and that fulfill the above conditions (i)-(iii).

The subtle points regarding CP of the dynamical maps stem from the possibility to drop out or at least relax some of the above assumptions (i)-(iii). The minimal contents of the complete positivity assumes validity of both equations (1) and (2), i.e. the condition (ii) of Definition 1. As emphasized above, such maps are called the completely positive and trace preserving  maps. The most often regarded and used maps in the field of open quantum systems are described by Definition 1. Dropping out the condition (iii) while maintaining the points (i) and (ii) of Definition 1 introduces the subtleties [25-28] that we examine in Section 5. Dropping out the condition (i) of Definition 1 is specific for the field of quantum information and computation, e.g. [32], that here will not be elaborated.

\bigskip

\noindent {\bf 3 Quantum bipartitions}

\bigskip

\noindent {\bf Definition 2} Every split of a composite system $C$ into a {\it pair} of subsystems is called a {\it bipartition} of the $C$ system.

Typically, bipartition of a composite system is not unique, e.g. [5, 6, 30, 31]. Of interest for our considerations is the above-defined composite system $C=S+E$,
where the open system $S$ may also be a composite system. We are particularly interested in a bipartition of the open system $S$ into a pair of subsystems denoted $1$ and $2$ ($S=1+2$).

While the total composite system, $C$, is assumed to be isolated and therefore subject of the unitary (reversible) Schr\" odinger dynamics, it can be differently bipartitioned.
That is, the tripartite system $C=1+2+E$ may be bipartitioned as $C=1+(2+E)\equiv1+E'$  or $C=2+(1+E)\equiv2+E''$. The original environment $E$ may be in contact (interaction) with both $1$ and $2$ subsystems or with only one of them; needless to say, noniteraction of the $S$ system with $E$ is trivial in our context.

Then the total system's Hilbert state space, $\mathcal{H}_C$, can be differently tensor-factorized:

\begin{equation}
\mathcal{H}_C = \mathcal{H}_1 \otimes \mathcal{H}_2 \otimes \mathcal{H}_E = \mathcal{H}_S \otimes \mathcal{H}_E = \mathcal{H}_1 \otimes \mathcal{H}_{E'} = \mathcal{H}_2 \otimes \mathcal{H}_{E''},
\end{equation}

\noindent where one tensor-factorization represents one possible partition (one possible structure) of the composite system.

It is well known that different bipartitions of a composite system give rise to different amounts of correlations (quantum or classical) in the system's decompositions  [5, 6, 30, 31] (and the references therein). That is, the correlations present in a quantum state (pure or mixed) are {\it not} invariants of the change of bipartition (of the composite system's structure).
Intuitively, correlations in a composite system do not regard the system or any of its states, but regard the system's structure.

For completeness, here we give one possible formulation of the Pechukas' theorem [23, 33]:

\noindent (PT) In order for a dynamical map $\Phi_{(t,t_{\circ})}$ has the {\it whole} Banach space $\mathcal{B}_S$ in its domain, the initial tensor-product state

\begin{equation}
\rho_{C}(0) = \rho_S(0) \otimes \rho_E(0),
\end{equation}

\noindent for the closed $C=S+E$ system is a {\it necessary condition}; $\rho_E(0)$ is common for all the initial states $\rho_S(0)$ of the open $S$ system.

The Pechukas' theorem applies universally, for every possible bipartition of a composite system, e.g. eq.(3), and directly regards the above condition (iii), Definition 1. Bearing in mind, that a tensor-product form of a quantum state for one bipartition typically implies non-tensor-product form of the state for virtually any other bipartition of the system [5, 6, 30, 31], it is expectable that, e.g., a CP dynamics for the subsystem $1$ may lead to a non-CP dynamics for the $2$ subsystem and {\it vice versa}. In the next section we provide the general conditions for the CP dynamics for both $1$ and $2$ subsystems of the $S$ composite open system, which is  subject of a CP dynamics,  Definition 1.

Technically, the contents of the next section may be regarded a reminiscence of the more general considerations [32, 34-36].
Nevertheless, our considerations give rise to a conceptually clear and technically simple {\it algorithm}--not yet known for the more general cases--for derivation of the Kraus forms for the subsystems CP dynamics,
while making direct link to the subtleties of the concept of complete positivity--in which regard  the general considerations are rather non-transparent.

\bigskip

\noindent {\bf 4 Considerations of the subsystems' dynamics}

\bigskip

\noindent {\bf Theorem 1} For a composite open system $S$, which is subject of a completely positive dynamics, Definition 1, the initial tensor-product state for the bipartition $S=1+2$
is {\it necessary and sufficient} for complete positivity of the dynamics of {\it both} subsystems, $1$ and $2$.

\noindent Proof. The necessary condition comes directly from the Pechukas' theorem. Applying (PT) to the subsystem $1$ implies the initial tensor product state:

\begin{equation}
\rho_{C}(0) = \rho_1(0) \otimes \rho_{E'}(0),
\end{equation}

\noindent and analogously for the subsystem $2$ the initial state:

\begin{equation}
\rho_{C}(0) = \rho_2(0) \otimes \rho_{E''}(0).
\end{equation}

According to (PT), the state $\rho_{E'}(0)$ is the same for all $\rho_{1}(0)$, and also the state $\rho_{E''}(0)$ is the same for all $\rho_2(0)$. Then it is easy to see that the {\it only} initial state
that may fulfill the requirements (4)-(6) is the full tensor-product state:

\begin{equation}
\rho_{C}(0) = \rho_1(0) \otimes \rho_2(0) \otimes \rho_E(0).
\end{equation}

That the condition eq.(7) is  {\it sufficient} for complete positivity (Definition 1) of the dynamics of both subsystems $1$ and $2$, can be seen as follows.

Introduce an orthonormal basis of Hermitian operators, $\{g_{1i}\}$, acting on the Hilbert state-space for the subsystem $1$, and analogously a basis $\{h_{2j}\}$
for the subsystem $2$. This procedure is straightforward for the finite-dimensional $1$ and $2$ systems, which we are mainly concerned with;  the orthonormalization rule for the set of operators is chosen as $tr_1 (g_{1i} g_{1i'})=\delta_{ii'}$, where
$\delta_{ii'}$ standing for the ''Kronecker delta'', and analogously for the subsystem $2$. Then every Kraus operator $K_k(t)$ in eq.(1) can be presented as:

\begin{equation}
K_k(t) = \sum_{i,j} c^k_{ij}(t) g_{1i}\otimes h_{2j};
\end{equation}

\noindent the normalization rule gives $c^k_{ij}=tr(K_k(t) g_{1i}\otimes h_{2j})$.

Substitute eqs. (7) and (8) into eq.(1) and take the trace over the subsystem $2$. Then, in accordance with the above point (iii), without imposing any restrictions on the initial states $\rho_1(0)$ and $\rho_2(0)$, it directly follows:

\begin{equation}
\rho_1(t)=tr_2 \rho_{C}(t) = \sum_{i,i'} \left(\sum_{k,j,j'}  c^k_{ij}(t) c^{k\ast}_{i'j'}(t) tr_2 (h_{2j}\rho_2(0)h^{\dag}_{2j'})\right) g_{1i} \rho_1(0) g^{\dag}_{1i'}\equiv  \sum_{i,i'} b_{ii'}(t) g_{1i} \rho_1(0) g^{\dag}_{1i'},
\end{equation}

\noindent which satisfies the above point (i), while eq.(2) implies $tr_1\rho_1(t) = 1, \forall{t}$, which leads to $\sum_{i,i'} b_{ii'}(t) g_{1i'}^{\dag} g_{1i} = I_1$, that is to satisfied the condition (ii).

As we show below, the matrix $(b_{ii'}(t))$--well-defined for the finite dimensional subsystems $1$ and $2$--is Hermitian and semidefinite positive. That is, the matrix $(b_{ii'}(t))$ has non-negative (real) eigenvalues, $b_p$,
for every $t$. Then it can be diagonalized (in general, separately for every instant of time $t$):

\begin{equation}
b_{ii'} = \sum_p b_p u_{ip} u^{\ast}_{i'p}
\end{equation}

\noindent where $(u_{ip})$ is a unitary matrix.

Placing eq.(10) into eq.(9) easily gives rise to a Kraus form for the subsystem $1$:

\begin{equation}
\rho_1(t) = \sum_p K_{1p}(t) \rho_1(0) K^{\dag}_{1p}(t),
\end{equation}

\noindent with the subsystem's Kraus operators:

\begin{equation}
K_{1p}(t) = \sum_i\sqrt{b_p(t)} u_{ip}(t) g_{1i},
\end{equation}

\noindent which satisfy the completeness relation eq.(2), i.e. $\sum_p K^{\dag}_{1p}(t) K_{1p}(t) =\sum_{i,i'} b_{ii'}(t) g_{1i'}^{\dag} g_{1i}= I_1, \forall{t}$, that is the condition (ii) for the subsystem's dynamics. Everything analogously for the subsystem $2$.

Finally, we prove that the matrix $(b_{ii'}(t))$ is positive semidefinite for every $t$. From eq.(9):

\begin{equation}
b_{ii'}(t)= \sum_{k,j,j'}  c^k_{ij}(t) c^{k\ast}_{i'j'}(t) tr_2 (h_{2j}\rho_2(0)h^{\dag}_{2j'})\equiv tr_2 \sum_k D^k_{2i}(t) \rho_2(0) D^{k\dag}_{2i'}(t),
\end{equation}

\noindent where $D^k_{2i}(t) \equiv \sum_j c^k_{ij}(t) h_{2j}$.

For every instant of time $t$: it is obvious that $b_{ii'}=b^{\ast}_{i'i}$, which is the condition of Hermiticity of the matrix $B = (b_{ii'}(t))$, while the semidefinite positivity of the matrix $B $ means

\begin{equation}
v^{\dag}Bv = \sum_{i,i'} v_i^{\ast} b_{ii'} v_{i'} \ge 0, \forall { t},
\end{equation}

\noindent for \textit{ every } vector $v=(v_i)$. Placing eq.(13) into eq.(14) easily gives:

\begin{equation}
v^{\dag}Bv = tr_2 \sum_k A_{2k} \rho_2(0) A^{\dag}_{2k} =  \sum_k tr_2 \left(A_{2k} \rho_2(0) A^{\dag}_{2k}\right) \ge 0,
\end{equation}

\noindent where $A_{2k} \equiv \sum_{i} v_i^{\ast} D^k_{2i}$, and the inequality follows from the obvious semidefinite positiveness  $tr_2A_{2k} \rho_2(0) A^{\dag}_{2k}\ge 0$ for every $k$
and for every instant of time $t$. This completes the sufficient condition for the subsystems CP dynamics. Q.E.D.

The proof of Theorem 1 is constructive in that it directly provides an \textit{ algorithm} for obtaining the Kraus form eq.(11) for the $1$ subsystem from the composite system's Kraus form eq.(1): (i) Calculate the coefficients $c^k_{ij}(t)= tr (K_k(t) g_{1i}\otimes h_{2j})$ from eq.(8); (ii)
according to eq.(13) calculate the matrix entries $b_{ii'}(t)$; (iii) diagonalize the matrix $(b_{ii'}(t))$; (iv) The Kraus operators  directly follow from the substitution of the results into eq.(12).

Now it is straightforward to generalize eq.(7) to subsystems
of a three- or more-partite (finite-dimensional) open system
 $C$.
In the full analogy with the proof of eq.(7), it readily follows,
that the only initial state $\rho_S(0)$ allowing for CP for {\it every} subsystem's dynamics is tensor
product, $\rho_S(0) = \otimes_i \rho_i(0)$. Accordingly, it  also straightforwardly follows a generalization of the above described algorithm
for obtaining the Kraus form of a subsystem's CP dynamics.

\bigskip

\noindent {\bf 5 Reduced-domain complete positivity}

\bigskip

\noindent Due to (PT), the algorithm of the previous section breaks for the initial state $\rho_{S}(0)$ carrying  correlations for the subsystems $1$ and $2$. To see how the procedure may break, let us assume a separable (non-entangled)
initial state $\rho_{S}(0)= \sum_m p_m \sigma_{1m}\otimes \sigma_{2m}\neq \rho_1\otimes\rho_2$; $\sum_m p_m=1$, while the $\sigma_m$s are of the unit trace, $tr \sigma_m=1,\forall{m}$.
Substituting this initial state into eq.(1) with the use of eq.(8) readily gives, instead of eq.(9):

\begin{equation}
\rho_1(t)= \sum_m p_m \left( \sum_{i,i'} b^m_{ii'}(t) g_{1i} \sigma_{1m} g^{\dag}_{1i'}\right),
\end{equation}

\noindent where $b^m_{ii'}(t) = \sum_{k,j,j'} c^k_{ij}(t) c^{k\ast}_{i'j'}(t) tr_2 (h_{2j} \sigma_{2m} h^{\dag}_{2j'})$.

Then  diagonalization as in eq.(10) can be separately applied for every $m$ in eq.(16) giving rise to

\begin{equation}
\rho_1(t) = \sum_m p_m \left(\sum_p K^m_{1p}(t) \sigma_{1m} K^{m\dag}_{1p}(t)\right).
\end{equation}

It is possibly obvious that eq.(17) \textit{ cannot} be written in the form of eq.(11), which, in turn, is \textit{ required} for CP.
To fulfill the Kraus form eq.(11), in eq.(17) should appear unique initial state
 $\rho_1(0)$, which is now defined as
$\rho_1(0) = tr_2 \rho_{S}(0) = \sum_m p_m \sigma_{1m}$, with arbitrary values of $p_m$s satisfying the unit-trace (i.e. the normalization)
condition, $\sum_m p_m =1$. The trace preservation eq.(2), i.e. the condition (ii) of Definition 1, implies: $\sum_p K_{1p}^{m\dag}(t)K_{1p}^{m}(t)=I, \forall{m, t}$.

 As an illustration, let us consider a tripartite $C=1+2+3$
open system, whose dynamics
is completely positive (Definition 1). Let us assume the initial state of the form $\rho_C(0) = \rho_{12}(0) \otimes \rho_3(0)$ for the bipartition $(1+2)+3$, where $\rho_{12}(0)$ carries some correlations, i.e. is not of the tensor-product form. Then obviously $\rho_C(0) \neq \rho_1(0) \otimes \rho_{23}(0)$ as well as $\rho_C(0) \neq \rho_2(0) \otimes \rho_{13}(0)$ for the structures $1+(2+3)$ and $2+(1+3)$, respectively. Due to Theorem 1, the presence of the initial correlations for these structures of the composite $C$ system makes the dynamics
non-CP  for the corresponding subsystems: $1$, $2$, $2+3$ and $1+3$.
However, due to the above assumption for the $(1+2)+3$ structure, Theorem 1 implies CP dynamics for both subsystems $1+2$ and $3$.
Therefore complete positivity, Definition 1, does not regard the open system or any of its states  but the open system's structure.

The point strongly to be emphasized (an extension of the told in Section 2): for every  structure of the possibly composite open system, a map that is {\it not} CP in the sense of Definition 1 may still be completely positive for certain variations of Definition 1.
Notably, dropping out the condition (iii) of Definition 1 may give rise to complete positivity on the reduced domain in the open-system's state space. Then the map is completely positive, i.e. can be presented in the form of eq.(1) (while maintaining the conditions (i) and (ii) of Definition 1), only for certain {\it special}  initial states of the open system.  This delicate point [25-28] can now be easily illustrated.

If reduced to a special set of states $\sigma_{1m}$ in eq.(16), equation (17) may in principle take the
form of eq.(11) and hence exhibit the domain-dependent complete positivity. As an example, let us assume that, for the map implicit to eq.(16), there exist some  $\sigma_{1m}$s, $m=1,2$, {\it such} that
 $K^m_{1p}$s are the same for $m=1,2$,
i.e. $K^1_{1p}=K^2_{1p}=K_{1p}$, for every index $p$ and every instant of time $t$. Then the right hand side of eq.(17) \textit{reduced} to only $m=1,2$ gives:

\begin{equation}
\sum_p K_{1p} \left(\sum_{m=1}^2 p_m \sigma_{1m}\right) K^{\dag}_{1p}.
\end{equation}

\noindent Bearing in mind that reduction to only $m=1,2$ gives for the $1$ system's initial state $\sigma_1(0)=tr_2\rho_{C}(0)=\sum_{m=1}^2 p_m \sigma_{1m}$, we can see that eq.(18) is of the Kraus form of eq.(1), i.e. that it gives
completely positive dynamics  for the \textit{reduced} domain of the initial states, $\sigma_{1m}$, for \textit{arbitrary} $p_m$s that satisfy the normalization condition  $\sum_{m=1}^2 p_m=1$. For every initial state not representable by $\sigma_1(0)=\sum_{m=1}^2 p_m \sigma_{1m}$, the map cannot take the form of eq.(1), i.e. of eq.(11).

Equations (16)-(18) and the thereof conclusions equally regard the alternative situations, when the subsystem $2$ does not exist (non-composite open system $C$) as well as the case when the composite system $C$ is built from the open system $1$ and some system $2$, which is a part of the environment.

\bigskip

\noindent {\bf 6 Application: a case study}

\bigskip

\noindent As an illustration of the general results of the previous sections, we consider a two-qubit system in the weak interaction with a thermal environment of mutually non-interacting harmonic oscillators (or ''normal modes'').
For completeness, we describe the physical background, but uninterested reader may skip  to eq.(23).

The total, isolated, system is described by the Hamiltonian [29]:

\begin{equation}
H = H_{1\circ} + H_{2\circ} + H_{E_1\circ} + H_{12} + H_{1E_1} ,
\end{equation}

\noindent where the index "$\circ"$ stands for the subsystems' self-Hamiltonians and the rest are the interaction terms.
Hence only the qubit $1$ is monitored by its environment denoted $E_1$.

While the self-Hamiltonians  are standard (see below), the qubits interaction
is chosen [29]:

\begin{equation}
H_{12} = \beta S_{1z} \otimes S_{2z},
\end{equation}

\noindent where the 1/2-spin operators $S_{p  z} = \sigma_{p z} /2, p = 1,2$ and we take $\hbar =1$, while the interaction with the environment:

\begin{equation}
H_{1 E_{1}} =S_{1x} \otimes \int_0^{\nu_{max}} d\nu h(\nu) (a^{\dag}_{\nu} + a_{\nu} ) \equiv  S_{1x} \otimes B_{E_1},
\end{equation}

\noindent where appear the annihilation and creation operators satisfying the standard Bose-Einstein commutation $[a_{\nu},a^{\dag}_{\nu'}]
=  \delta (\nu - \nu') $.

The total system's self-Hamiltonian [in the units of $\hbar=1$]

\begin{equation}
H_{\circ} = H_{1\circ} + H_{2\circ}  +  H_{12} + H_{E_1\circ}   = {\omega \over 2} \sigma_{1z} + {\omega \over 2} \sigma_{2z} + {\beta \over 4} \sigma_{1z} \otimes \sigma_{2z} + H_{E_1\circ},
\end{equation}

\noindent where the environmental self-Hamiltonian: $H_{E_1\circ} =  \int_0^{\nu_{max}} d\nu a^{\dag}_{\nu} a_{\nu}$ with the maximum cutoff frequency $\nu_{max}$.
Initial state of the environment is assumed to be thermal,  and the total system's initial state is tensor-product, $\rho(0)=\rho_{C}\otimes\rho_{E_1}$; $C=1+2$.

The following set of the Hermitian Kraus operators for the pair of qubits is found in the interaction picture [29]:

\begin{equation}
K_1={\sqrt{1-e^{-32t\gamma_2}}\over 2}
\left(\begin{array}{cccc}
0&0&0 &0\\
0&0&0 &\imath\\
0&0&0 &0\\
0&-\imath &0 &0
\end{array}\right),
K_2={\sqrt{1-e^{-32t\gamma_2}}\over 2}
\left(\begin{array}{cccc}
0&0&0 &0\\
0&0&0 &-1\\
0&0&0 &0\\
0&-1 &0 &0
\end{array}\right),
\end{equation}

\begin{equation}
K_3={\sqrt{1-e^{-32t\gamma_1}}\over 2}
\left(\begin{array}{cccc}
0&0&1 &0\\
0&0&0 &0\\
1&0&0 &0\\
0&0 &0 &0
\end{array}\right),
K_4={\sqrt{1-e^{-32t\gamma_1}}\over 2}
\left(\begin{array}{cccc}
0&0&-\imath &0\\
0&0&0& 0\\
\imath&0&0 &0\\
0&0 &0 &0
\end{array}\right),
\end{equation}

\begin{equation}
K_5={1-e^{-16t\gamma_2}\over 2}
\left(\begin{array}{cccc}
0&0&0 &0\\
0&-1&0 &0\\
0&0&0 &0\\
0&0 &0 &1
\end{array}\right),
K_6={1-e^{-16t\gamma_1}\over 2}
\left(\begin{array}{cccc}
1&0&0 &0\\
0&0&0 &0\\
0&0&-1 &0\\
0&0 &0 &0
\end{array}\right).
\end{equation}

\noindent The damping functions are given by expressions:
\begin{eqnarray}
&\nonumber&
\gamma_1\equiv 4\pi J(\omega+\beta/2)\bar{n}(\omega+\beta/2) , \nonumber
\\&&
\gamma_2\equiv 4\pi J(\omega-\beta/2)\bar{n}(\omega-\beta/2) ,
\end{eqnarray}
 in the high-temperature limit.  $\bar{n}(\nu)$ is  the average number of bosons in thermal state $\bar{n}(\nu)=
(e^{-\nu/T}-1)^{-1}$ while $J(\nu)=\alpha\nu e^{-\nu/\nu_c}$ stands for  the standard Ohmic spectral density with the cutoff $\nu_c$.

The  last two Kraus matrices are diagonal and rather large. So we present their non-zero entries with the use of the following notation:
 $\tau = (\gamma_1+\gamma_2)t$,
$W=(\gamma_1-\gamma_2)/(\gamma_1+\gamma_2)$.

For $K_7$: $K^7_{1,1}=K^7_{3,3}=A
(-8e^{32\tau}+2e^{24\tau}\sinh(16W\tau) + 4e^{32\tau}\sinh(8W\tau)+B)$;
$K^7_{2,2} = K^7_{4,4} = A(-8e^{32\tau}-2e^{24\tau}\sinh(16W\tau) - 4e^{32\tau}\sinh(8W\tau)+B)$.

For the $K_8$ matrix:
$K^8_{1,1}=K^8_{3,3}=-A'(8e^{32\tau}-2e^{24\tau}\sinh(16W\tau) - 4e^{32\tau}\sinh(8W\tau)+B)$;
$K^8_{2,2}=K^8_{4,4}=-A'(8e^{32\tau}+2e^{24\tau}\sinh(16W\tau) + 4e^{32\tau}\sinh(8W\tau)+B)$.

In $K_7$ and $K_8$ appear:

\begin{equation}
A=\sqrt{2e^{-16\tau} + 2e^{-32\tau}\cosh(16W\tau) + 4e^{-24\tau} \cosh(8W\tau) - e^{-56\tau}B
 \over
 16(2e^{16\tau}\sinh(16W\tau)+4e^{24\tau}\sinh(8W\tau))^2 + 16e^{-16\tau}(B-8e^{32\tau})^2 }
\end{equation}

\begin{equation}
A'=\sqrt{2e^{-16\tau} + 2e^{-32\tau}\cosh(16W\tau) + 4e^{-24\tau} \cosh(8W\tau) + e^{-56\tau}B
 \over
 16(2e^{16\tau}\sinh(16W\tau)+4e^{24\tau}\sinh(8W\tau))^2 + 16e^{-16\tau}(B+8e^{32\tau})^2 }
\end{equation}

\noindent and

\begin{eqnarray}
&\nonumber& B^2=
2e^{48\tau}(28e^{16\tau}-1) + 2e^{48\tau} \cosh(32W\tau) - 8e^{56\tau}\cosh(8W\tau)
\\&&
+ 8e^{64\tau} \cosh(16W\tau) + 8e^{280\tau} \cosh(120W\tau).
\end{eqnarray}

\noindent These Kraus operators, $K_i(t), i=1,2,...,8$, should be 	
 placed in eq.(1) for the pair of qubits $1+2$ modelled by eqs.(19)-(22).
The Kraus operators in the Schr\" odinger picture are defined as $\exp(-\imath H^{C}_{\circ}t) K_i(t) $,
where $H^{C}_{\circ}$ is the $H_{\circ}$ term in eq.(22) \textit{with dropped} the $H_{E_1\circ}$ term.

We are now ready to apply the algorithm of Section 4
regarding the qubit $1$; everything analogous for the qubit $2$. According to Section 4, we assume a tensor-product initial state $\rho_{C}(0) = \rho_1(0)\otimes \rho_2(0)$, {\it without imposing any restrictions} to the choices of $\rho_1(0)$ and $ \rho_2(0)$.
What follows is typical for the calculations of the sort:
plenty of straightforward but lengthy algebraic calculations. Therefore we only provide the main steps of the calculation that can be patiently and straightforwardly reproduced.

As an orthonormal basis of the operators $g_{1i}\otimes h_{2j}$
with the orthonormalization rule $tr (g_{1i}\otimes h_{2j} g_{1i'}\otimes h_{2j'}) = \left(tr_1(g_{1i}g_{2i'})\right) \left(tr_2(h_{2j}h_{2j'})\right) = \delta_{ii'} \delta_{jj'}$ (see Section 4)
for the pair of qubits, we choose the standard, so-called Pauli basis, $\sigma_{1i}\otimes\sigma_{2j}/2, i,j=0,1,2,3$, where for both qubits, $\sigma_{\circ}=I$
and $\sigma_i, i=1,2,3$, are the standard Pauli matrices; the single-qubit basis, $\{\sigma_{i}/\sqrt{2}, i=0,1,2,3\}$.

To reduce the number of independent parameters, due to equations (23)-(29), for the complex elements, we place $K^{i\ast}_{qp}=-K^i_{pq}$, where $K^i_{pq}$ is the $(pq)$th (the $p$th row and the $q$th column) element of the $K_i$ Kraus matrix. For example, non-zero $K^1_{24}= K^{1\ast}_{42} =\imath\sqrt{1-\exp(-32t\gamma_2)}/2$,  etc. Then the item (i) of the above algorithm easily leads to the following set of the non-zero $c$-parameters defined by eq.(8): $c^1_{20}=-c^1_{23}=\imath K^1_{24};
c^2_{10}=-c^2_{13}=K^2_{24};
c^3_{10}=  c^3_{13} = K^3_{13};
c^4_{20}=c^4_{23}=-\imath K^4_{31};
c^5_{30} = -c^5_{33}=K^5_{22};
c^6_{30}=c^6_{33}=K^6_{11};
c^7_{00}= K^7_{11} + K^7_{22}, c^7_{03}= K^7_{11} - K^7_{22};
c^8_{00}= K^8_{11} + K^8_{22}, c^8_{03}= K^8_{11} - K^8_{22}
$.

The item (ii) of the algorithm returns a diagonal matrix $B$, universally; this makes the item (iii) of the algorithm unnecessary.
The entries of the matrix $B$, of course, depend on the initial state of the qubit $2$, $\rho_2(0)$. Without loss of generality, we assume \textit{arbitrary} initial pure
state $\sqrt{a}\vert +\rangle_2 + \sqrt{1-a}\vert -\rangle_2$ (where $\sigma_{2z}\vert\pm\rangle_2 = \pm\vert\pm\rangle_2$ and, for simplicity, $a$ is real, $0\le a\le 1$), for which it follows:

\begin{eqnarray}
&\nonumber& b_0 \equiv b_{00} = 2a\vert K^7_{11} \vert^2 + 2(1-a)\vert  K^7_{22} \vert^2 + 2a\vert K^8_{11} \vert^2 + 2(1-a)\vert  K^8_{22} \vert^2
 \\&&\nonumber b_1\equiv b_{11} = 2(1-a)\vert K^2_{24}\vert^2 + 2a\vert K^3_{13}\vert^2
\\&& \nonumber
 b_2\equiv b_{22} = 2(1-a)\vert K^1_{24}\vert^2 + 2a\vert K^4_{31}\vert^2
\\&& b_3\equiv b_{33} = 2(1-a)\vert K^5_{22}\vert^2 + 2a\vert K^6_{11}\vert^2.
\end{eqnarray}

The use of eq.(12), i.e. the item (iv) of the algorithm, directly leads to the Hermitian subsystem's Kraus operators in the interaction picture:

\begin{equation}
k_{\circ} =\sqrt{b_{\circ}\over 2} I, \quad k_1= \sqrt{b_1\over 2} \sigma_x, \quad k_2=\sqrt{b_2\over 2} \sigma_y, \quad k_3=\sqrt{b_3\over 2} \sigma_z.
\end{equation}

Substituting the matrix elements of the Kraus operators eq.(23)-(29) into eq.(30), a lengthy but straightforward calculation gives rise to the explicit time dependence:

\begin{eqnarray}
&\nonumber& b_0  ={1\over 2}+{1-a\over 2}e^{-32\gamma_2t}+{a\over 2}e^{-32\gamma_1t}+(1-a)e^{-16\gamma_2t}+ae^{-16\gamma_1t}
 \\&&\nonumber b_1= b_2= {1\over 2} - {1-a\over 2}e^{-32t\gamma_2}-{a\over 2}e^{-32t\gamma_1}
\\&& b_3= {1-a\over 2}\left(1-e^{-16t\gamma_2}\right)^2 +{a\over 2}\left(1-e^{-16t\gamma_1}\right)^2
\end{eqnarray}

\noindent that gives the following state of the qubit $1$ in the interaction picture:

\begin{equation}
\rho_1(t) = \Phi^1_{(t,0)}\rho_1(0) = {b_{\circ}(t)\over 2} \rho_1(0) + {b_1(t)\over 2} \sigma_x\rho_1(0)\sigma_x + {b_1(t)\over 2} \sigma_y\rho_1(0)\sigma_y + {b_3(t)\over 2} \sigma_z\rho_1(0)\sigma_z.
\end{equation}

Taking the trace of $\rho_1(t)$ gives the completeness condition eq.(2) satisfied:

\begin{equation}
tr_1 \rho_1(t) = \sum_i k_i(t) k_i(t) = {1\over 2} (b_{\circ}(t) + b_1(t) + b_2(t) + b_3(t)) = 1
\end{equation}

\noindent for every instant of time $t$. Therefore, the  qubit's dynamical map $\Phi^1_{(t,0)}$ is both of the Kraus form eq.(1) and trace preserving, as physically it should be, while no restriction to the initial state $\rho_1(0)$ has been imposed--in accordance with the condition (iii) of Definition 1.

It is worth stressing that the Kraus form eq.(11) exists independently of the strength of
the qubits mutual interaction. For different values of the interaction strength $\beta$, the Kraus operators are different, but the fact of their
existence is out of question. Dependence of the Kraus operators on the initial state of the subsystem $2$ is given by eqs.(30) and (31) for  arbitrary initially pure state $\rho_2(0)$, and analogously for the mixed states
while bearing in mind the general form  $\rho=(I + \vec{\sigma} \cdot \vec {n})/2$ of a state of a single qubit; for pure states, $\vert \vec n\vert = 1$ [3].
The variations of the initial state (i.e. of the parameter $a$) give rise to variations of the contributions of the damping rates $\gamma_{1,2}$ in eq.(26) without alternating anything else in eq.(33).
In passing, we note that the single-qubit dynamics presented by equations (32) and (33) is known to be Markovian [37].

Obtaining the Schr\" odinger-picture form of the Kraus operators eq.(31) is straightforward only for non-interacting qubits (when $\beta = 0$); then the Schr\" odinger picture operators read $\exp(-\imath t H_{1\circ})k_i(t)$.
In general, one should transform the total-system's Kraus operators into the Schr\" odinger picture, $\exp(-\imath H^{C}_{\circ}t) K_i(t) $ (see above), and then apply the algorithm.

\bigskip

\noindent {\bf 7 Discussion}

\bigskip

\noindent Section 4 directly concerns the finite-dimensional open systems. For the infinite-dimensional (''continuous variable'')
open systems, every step should be carefully checked if applicable.

A special case of our considerations is provided by the condition $c^k_{ij} = a^k_i b^k_j$ for eq.(8), which directly leads to the tensor product Kraus operators
$K_k = K_{1k}\otimes K_{2k}$, where $K_{1k}= \sum_i a^k_i g_{1i}$, and analogously for the subsystem $2$.
This requires the mutually non-interacting subsystems $1$ and $2$;
in the context of Section 6, this is the  case $\beta = 0$ in eq. (20). Nevertheless, unless \textit{all} the Kraus operators are already given in the tensor-product form,
the procedure of Section 4 should be applied. An alternative route may be taken by investigating whether or not  $c^k_{ij} = a^k_i b^k_j$ for \textit{every} $k$. To this end, the method developed in Ref. [38] may be useful.

Non-invariance of quantum correlations with respect to a change of structure of open system [5, 6, 30], that equally regards the quantum discord [39], sheds new light on the Pechukas' theorem: domain of a dynamical map regards the open system's structure by distinguishing those structures that can fulfill the condition eq.(4). Going beyond the considerations of Section 3, we can say that "structure"   regards the border line between the "open system" and "environment"; not only the subsystems of an open composite system may be of interest, but also the composite system {\it built} from the open system and a part of its environment [40]. Thus every redefinition of the "open system" changes conclusions obtained for the "original" open system.

Once properly linked with the Pechukas' theorem,
the concept of structure offers a well-suited background for describing the subtleties of the concept of complete positivity. On the one hand, it clearly exhibits the subtlety of the reduced domain of complete positivity, Section 5. On the other hand, it also provides a basis for designing an algorithm for constructing the subsystems' Kraus operators--if such exist (Theorem 1 of Section 4).
To this end, the more general considerations [32, 34-36] are rather non-transparent while not offering a clear and general algorithm for deriving the subsystems' Kraus operators, if such exist. Usefulness of the less general considerations (Definition 1 versus its variations) reminds us of the benefits of studying the POVM measurements despite the well defined general quantum measurements [3]. That is, the less general cases may be more transparent, even instrumental for plenty of the situations of interest in application. Definition 1 distinguishes an important class of the dynamical maps of interest in both open quantum systems theory as well as in the quantum information and computation science and is therefore the main object of our interest in this paper.

\bigskip

\noindent {\bf 8 Conclusion}

\bigskip

\noindent The subtleties of the concept of complete positivity can be conceptually clearly and technically simply presented by addressing the question of complete positivity of the subsystems' dynamics. In regard of this, we provide the conditions of existence and a simple algorithm for deriving the subsystems' Kraus operators. An example of a pair of mutually interacting qubits is illustrative and instrumental for both the conceptual background as well as for application of the algorithm.

\bigskip

\noindent {\bf Acknowledgements} Work on this paper is financially supported by Ministry of Science Serbia, grant 171028, and in part for MD by the ICTP -- SEENET-MTP project NT-03 Cosmology-Classical and Quantum Challenges.

\bigskip

\noindent {\bf References}

\bigskip

[1] Breuer, H.-P.,   Petruccione, F.: The Theory of Open Quantum Systems. Oxford University Press, New York (2002)

[2] Rivas, \' A., Huelga, S.F.: Open Quantum Systems. An Introduction. Springer, Berlin (2012)

[3] Nielsen, M.A., Chuang, I.L.: Quantum Computation and Quantum Information. Cambridge University Press, Cambridge (2000)

[4] Zurek, W.H.: Decoherence, einselection, and the quantum origins of the classical. Rev, Mod. Phys. 75, 715-775 (2003)

[5] Dugi\' c, M., Jekni\' c, J.: What is "system": some  decoherence-theory arguments. Int. J. Theor. Phys. 45,  2215-2225 (2006)

[6] Dugi\' c, M., Jekni\' c-Dugi\' c, J.: What is "system": the information-theoretic arguments. Int. J. Theor. Phys. 47, 805-813 (2008)

[7] Dugi\' c, M.,  Jekni\' c-Dugi\' c, J.:  Parallel decoherence in composite quantum systems . Pramana 79, 199-209  (2012)

[8] Jekni\' c-Dugi\' c, J.,  Dugi\' c, M.,  Francom, A.:  Quantum Structures of a Model-Universe: An Inconsistency with Everett Interpretation of Quantum Mechanics.  Int. J. Theor. Phys.  53,  1483-1494 (2014)

[9] Walls, D.F.,  Milburn, G.J.: Quantum Optics. Springer, Berlin (1994)

[10] Diosi, L.:  Gravity-related spontaneous collapse in bulk matter. New. J. Phys. 16, 105006 (2014)

[11] Nelson, E.,  Jess Riedel, C.: Classical branches and entanglement structure in the wavefunction of cosmological fluctuations, E-print  arXiv:1711.05719v1 [quant-ph]

[12] Schleich, W.P., Ranade, K.S., Anton, C.: Quantum technology: from research to application. Appl. Phys. B May 2016, 122-130 (2016).

[13] Biele, R.,  Rodriguez-Rosario, C.A.,  Frauenheim, T.,  Rubio, T.: Controlling heat and particle currents in nanodevices by quantum observation.
npj Quantum Materials 2, 38 (2017)

[14] Kassal, I.,  Yuen-Zhou, J., Rahimi-Keshari, S.: Does Coherence Enhance Transport in Photosynthesis?. J. Phys. Chem. Lett. 4, 362-367 (2013)

[15] Chen, H.-B., Lambert, N., Cheng, Y.-C., Chen, Y.-N., Nori, F.: Using non-Markovian measures to evaluate quantum master equations for photosynthesis.  Sci. Reports 5, 12753 (2015)

[16] Kottas, G.S., Clarke, L. I., Horinek D., Michl J.: Artificial Molecular Rotors.   J. Chem. Rev. 105, 1281-1376 (2005)

[17] Jekni\' c-Dugi\' c, J., Petrovi\' c, I.,  Arsenijevi\' c, M., Dugic, M.: Dynamical stability of the one-dimensional rigid
Brownian rotator: The role of the rotator's spatial size and shape. J. Phys. Cond. Matt. (2018). https://doi.org/10.1088/1361-648X/aab9ef

[18] Gemmer, J., Michel, M., Mahler, G.:  Quantum Thermodynamics. Emergence of Thermodynamic Behavior Within Composite Quantum Systems.     Springer-Verlag, Berlin  (2009)

[19] Taniguchi, N.: Quantum thermodynamics of nanoscale steady states far from equilibrium. Phys. Rev. B 97, 155404 (2018)

[20] Kraus, K.: States, effects and operations, fundamental notions of quantum theory. Springer-Verlag, Berlin (1983)

[21] Rivas, \' A., Huelga, S.F.,  Plenio, M.B.: Quantum Non-Markovianity: Characterization, Quantification and Detection. Rep. Prog. Phys. 77, 094001 (2014)

[22] Breuer, H.-P., Laine, E.-M., Piilo, J.,  Vacchini, B.: Non-Markovian dynamics in open quantum systems. Rev. Mod. Phys.  88, 021002 (2016)

[23] Pechukas, P.: Reduced Dynamics Need Not Be Completely Positive. Phys. Rev. Lett. 73, 1060-1062 (1994)

[24] Ferialdi, L.: Dissipation in the Caldeira-Leggett model. Phys. Rev. A 95, 052109 (2017)

[25]  Shabani, A., Lidar, D.A.: Vanishing Quantum Discord is Necessary and Sufficient for Completely Positive Maps.
Phys. Rev. Lett. 102, 100402 (2009)

[26] Brodutch, A., Datta, A.,  Modi, K.,  Rivas, \' A., Rodriguez-Rosario, C.A.: Vanishing quantum discord is not necessary for completely positive maps. Phys. Rev. A 87, 042301 (2013)

[27] Shabani, A.,  Lidar, D.A.: Erratum: Vanishing Quantum Discord is Necessary
and Sufficient for Completely Positive Maps [Phys. Rev. Lett. 102, 100402 (2009)]. Phys. Rev. Lett. 116, 049901 (2016)

[28] Kumar Sabapathy, Solomon Ivan, K.J.,  Ghosh, S.,   Simon, R.: Quantum discord plays no distinguished role in characterization of complete positivity: Robustness of the traditional scheme.
E-print arXiv:1304.4857 (2013)

[29] Arsenijevi\' c, M.,   Jekni\' c-Dugi\' c, J.,   Dugi\' c, M.: Kraus operators for a pair of interacting qubits: a case study.
E-print  	arXiv:1708.04172 [quant-ph].

[30] Jekni\' c-Dugi\' c, J.,  Arsenijevi\' c, M.,  Dugi\' c , M.: Quantum Structures. A View of the Quantum World. LAP Lambert Academic Publishing, Saarbr\" ucken  (2013)

[31]
 Quantum Structural Studies. Classical Emergence from the Quantum Level. eds. R. E. Kastner, J. Jekni\' c-Dugi\' c, G. Jaroszkiewicz. World Scientific Publishing, Singapore (2017)

[32] Lu, X.-M.: Structure of correlated initial states that guarantee completely positive reduced dynamics.  Phys. Rev. A  9, 042332 (2016)

[33] Jordan, T.F.,  Shaji, A.,  Sudarshan, E.C.G.: Dynamics of initially entangled open quantum systems. Phys. Rev. A 70, 052110 (2004)

[34] Buscemi, F.: On complete positivity, Markovianity, and the quantum data-processing inequality, in the presence of initial system-environment correlations.  Phys. Rev. Lett. 113, 140502 (2014)

[35] Dominy, J.M., Shabani, A.,  Lidar, D.A.: A general framework for complete positivity. Quantum Inf. Process.  15, 465-494 (2016)

[36] Vacchini, B., Amato, G.:  Reduced dynamical maps in the presence of initial correlations. Sci. Rep.  6, 37328 (2016)

[37]  Andersson, E.,  Cresser, J.D.,  Hall, M.J.W.: Finding the Kraus decomposition from a master equation and vice versa. J. Mod. Opt. 54, 1695-1716 (2007)

[38] Dugi\' c, M.: On diagonalization of a composite-system observable. Separability. Phys. Scripta 56, 560-565 (1997)

[39] Dugi\' c, M., Arsenijevi\' c, M., Jekni\' c-Dugi\' c, J.: Quantum correlations relativity for continuous variable systems.
Sci. China Phys. Mech. Astronomy  56, 732-736 (2013)

[40] Arsenijevi\' c, M., Jekni\' c-Dugi\' J. ,  Todorovi\' c, D.,  Dugi\' c, M.: Entanglem,ent relativity innthe foundations of the open quantum systems theory.
In: Watson, L., (ed.) New Research on Quantum Entanglement, pp. 99-116. Nova Science Publishers, Hauppauge NY  (2015)

\end{document}